\documentclass[prl,aps,twocolumn,amsmath,amssymb,floatfix,showpacs]{revtex4}

\usepackage{graphicx}
\usepackage{dcolumn}
\usepackage{bm}

\begin{document}

\title{Is Small Perfect? Size Limit to Defect Formation in Pyramidal Pt Nanocontacts}

\author{V. Rodrigues$^{1}$}
 \email{varlei@ifi.unicamp.br}
\author{F. Sato$^{1}$}
\author{D. S. Galv\~{a}o$^{1}$}
\author{D. Ugarte$^{1,2}$}
 
\affiliation{$^{1}$Instituto de F\'{i}sica ``Gleb Wataghin'', Universidade Estadual de Campinas, Unicamp, C.P. 6165, 13083-970, Campinas, S\~{a}o Paulo, Brazil}

\affiliation{$^{2}$Laborat\'{o}rio Nacional de Luz S\'{\i}ncrotron, C.P. 6192,
13084-971, Campinas , S\~{a}o Paulo, Brazil}

\date{\today}

\begin{abstract}
We report high resolution transmission electron microscopy and \textit{ab initio} calculation results for the defect formation in Pt nanocontacts (NCs). Our results show that there is a size limit to the existence of twins (extended structural defects). Defects are always present but blocked away from the tip axes. The twins may act as scattering plane, influencing contact electron transmission for Pt NC at room temperature and Ag/Au NC at low temperature.
\end{abstract}


\maketitle

The structural and mechanical properties of nanometric wires represent a fundamental issue for the understanding of different phenomena such as friction, fracture, adhesion, etc. \cite{ref10}. There is a renewed interest in such systems motivated in part by the growing demand on miniaturization, new functionalities and less power consuming systems. Components such as diodes \cite{ref1}, switches \cite{ref2}, and electronic mixers \cite{ref3} have been built with simple molecules as active units \cite{ref4}. In spite of these technological advances, important issues on device integration, how to connect them in a stable and reproducible way, are still open \cite{ref5,ref6}. In order to build functional nanodevices, electrical contacts of nanometric dimensions are needed. 
The contact atomic arrangement will probably determine the electronic structure and, in consequence, the coupling between leads and the active part (for example, molecules) of the device \cite{ref6,ref7,ref8}. In this context, it is necessary to understand the influence of contact properties in the device characteristics and also the role played by structural defects since they can compromise the device functionality and reliability. Unfortunately, we can not control yet the nanocontact (NC) fabrication and, its characterization is rarely possible; consequently, control and reproducibility are still very difficult, even for the simplest cases, such as the hydrogen molecule \cite{ref9}.

Metal conductor NC studies have been usually done in experiments where the NCs are generated by mechanical elongation. 
In order to get a deeper understanding on the NC properties, its mechanical and atomic structure should be better comprehended. For example, for atomic size NC generated by mechanical stretching, we can expect that the metal deformation must exhibit stages where structural defects are formed to acomodate the elongation strain. 
It has been already shown that NCs obtained by mechanical elongation exhibit distinct elastic strain stages, followed by sharply defined yielding, originated from structural reorganization \cite{ref10,ref20, ref11}. In this way, we can not properly address the deformation of nanosystems using continuum-based theories, a microscopic analysis taking into account the atomic structure is necessary. Today, most of atomistic process information comes from molecular dynamics simulations where the elongation velocities (m/s) are orders of magnitude larger than in experiments (nm/s) \cite{ref10,ref12,ref13,ref14}. The plastic deformation is associated with collective slips of entire atomic planes or order-disorder transitions \cite{ref11,ref12}. It has been suggested that at very small scales (10\ nm) dislocations should be completely suppressed because the involved stress are comparable to the intrinsic lattice strength. In analogy, a self purification effect has been recently reported for semiconductor particles: defect formation energy increases as the nanocrystal size decreases, rendering difficult the doping of smaller systems \cite{ref16,ref17}.

In this sense some fundamental questions must be addressed. Is small perfect? Can we generate defect free metallic NCs? Is there a size limit to the existence of such structures? In this letter we present results from dynamic high resolution transmission electron microscopy (HRTEM) experiments of the atomic arrangement of Pt NCs under mechanical deformation at room temperature.
We have observed that there is a size limit where the twins (extended structural defects) can exist, below this limit the defects are always present but blocked few atomic planes away from the atomically sharp Pt NC. In order to quantitatively estimate the size limit where defect free contacts can exist, we have estimated the energetics associated with the twin generation through \textit{ab initio} calculations using the well-known and tested SIESTA code \cite{ref19}.

The atomic size metal NCs were generated \textit{in situ} in a HRTEM (JEM 3010 URP, 300\ kV, 0.17\ nm point resol., operating at LME/LNLS, Campinas, Brazil). In this methodology, the microscope electron beam is focused to generate holes in a self-supported thin metal film until neck is formed between them \cite{ref23}. The NCs (1-2\ nm in diameter) spontaneously become thinner due to displacements of the apexes, reach the size of a few atoms and finally break. During this process, we have no control on the direction or speed of the elongation. Real time evolutions were recorded using a high sensitivity TV camera (Gatan 622SC, 30 frame/s).

For the theoretical analysis, we have used \textit{ab initio} density functional theory (DFT) in the framework of the local density approximation (LDA) \cite{ref19}. In order to warranty high precision results, we have used double- basis set plus polarization functions (relativistic calculations) and norm-conserving pseudopotentials built on Troullier Martins scheme \cite{ref24}. We first tested the parameters used on SIESTA code for Pt perfect crystalline bulk structures, relaxing both cell lattice vectors and atomic positions. The obtained values are in good agreement with the experimental ones (3.924 and 3.875 \AA\ , for the experimental \cite{ref25} and SIESTA values, respectively). Once established the pseudopotential reliability for Pt structures, we proceeded to analyze the energetics and relative stability of the structures with defects (twins).

\begin{figure}[ht]
\begin{center}
\includegraphics[width=\columnwidth]{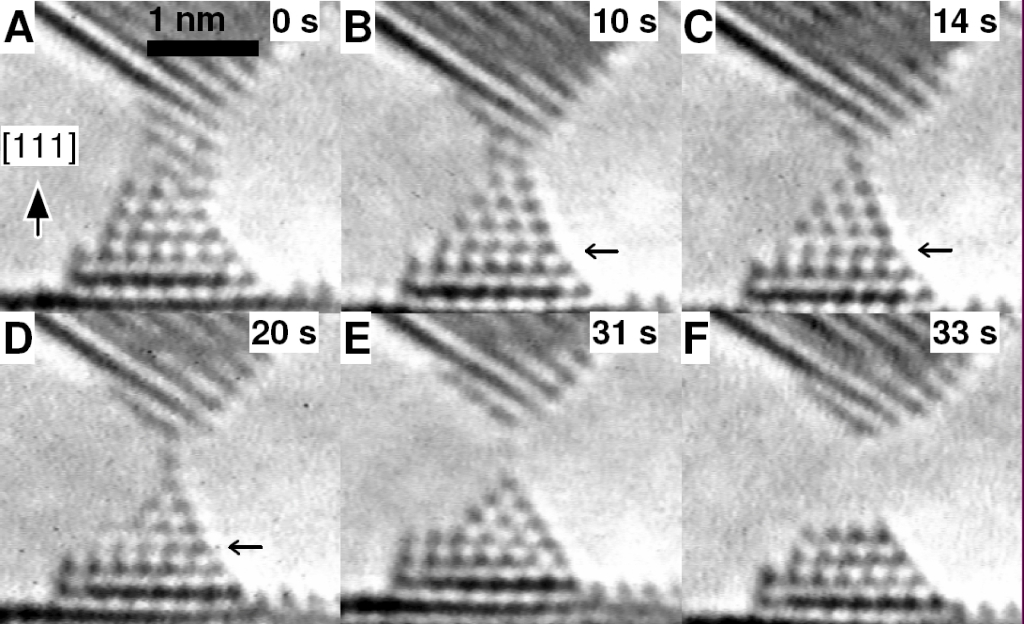}
\caption{Sequence of images showing the elongation and rupture of a Pt NC under tensile stress with a minor shear component (observation axis $[1\overline{1}0]$).}
\label{Pt_tip}
\end{center}
\end{figure}

The analysis of numerous HRTEM images always display defect free crystalline Pt nanoconstrictions composed of few atomic planes. When present, defects are observed away from the tip. 
As an illustrative example, figure \ref{Pt_tip} shows a
nanostructure where the lower apex is well aligned with the HRTEM beam, thus providing atomic and time resolved images of a pyramidal tip contact. Its axis is along the [111] crystalline direction (hereafter called [111] tip) and that ends with just only one atom. In the lower apex, the horizontal lattice periodicity is 0.24 $\pm$ 0.01 nm (Pt (111) spacing is 0.227 nm) and the angular relations confirm that the pyramid is defined by a (111) and (100) facet at left and right, respectively. The apex vertical movement generates the NC thinning, but a small shear movement (upper apex is gliding to the right) has induced the formation of a kink on the left apex side in (a). Further deformation generates an atomically sharp tip formed by 7 atomic layers (c); the arrow indicates the formation of a twin defect at the 5th atomic layer (counting from the tip); after rupture (f), the twin defect is annihilated and the tip reorganizes itself to form a truncated pyramid. See supplementary material in \cite{video1}.

\begin{figure}[ht]
\begin{center}
\includegraphics[width=\columnwidth]{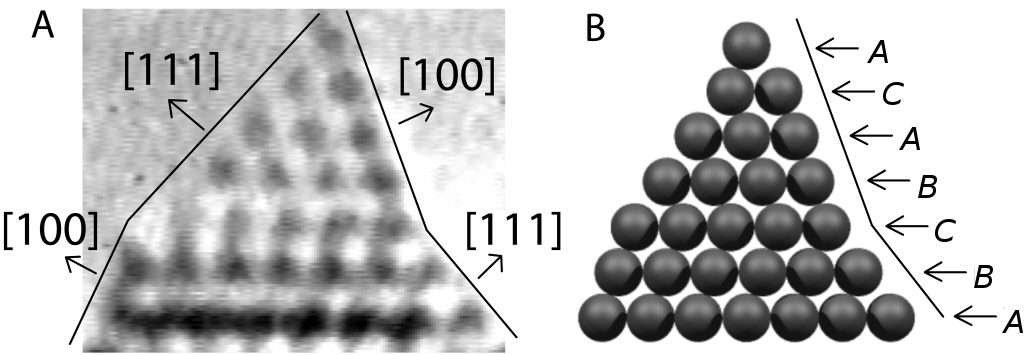}
\caption{(a) NC with a twin defect formed in its 5th layer counting from the top. (b) Schematic draw of the tip in (a) showing the usual $ABC$ notation for the face-centered-cubic packing.}
\label{twin}
\end{center}
\end{figure}

From figure \ref{Pt_tip} we can also observe the presence of planar defects, more precisely a twin defect. This can be more precisely identified by considering that a twin defect is a fault in the $ABC$ planes sequence forming a mirror: $\overleftarrow{ABC}\overrightarrow{BAC}$. On the experimental image it can be recognized by the angle change in the pyramidal facets. Above the twin, the left facet of the tip is a (111) plane and a (100) one on the right; bellow the twin it is a (111) plane on the right and (100) on the left (see figure \ref{twin}).

\begin{figure}[ht]
\begin{center}
\includegraphics[width=\columnwidth]{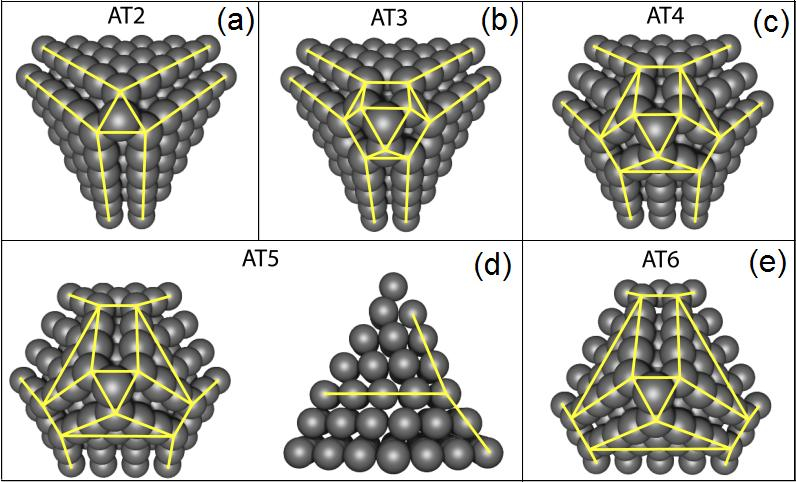}
\caption{Schematic draw of possible atomic arrangement of twinned [111] Pt NCs. The schemes in (a)-(e) are organized as a function of the twin height position at the apex, 2nd, 3rd, 4th, 5th and 6th layer respectively. Lines indicate the atomic layer and the facet borders.}
\label{modelo}
\end{center}
\end{figure}

One very important result is that during the NC formation no twin was observed very close to the tip. It has only been observed at the 4th layer (figure \ref{Pt_tip}(d)) and at the 5th layer (figure \ref{Pt_tip}(e)). These results raise the important question whether this represents the transition size between the macroscopic plasticity mechanism and the nanometric scale where extended defects can not be sustained anywhere \cite{ref10}. In order to better address this question, we have analyzed the energetics associated with this defect formation. We have calculated the structure relaxation and the total energy of the structures shown in figure \ref{modelo}(a)-(e). As the experimentally observed tip (figure \ref{Pt_tip}) has seven layers, we have investigated all the five possible positions for the twin defect, as shown in figure \ref{modelo}.

In order to model the tip based on our experimental data we should consider that the HRTEM images provide a bi-dimensional projection of the atomic arrangements, but tree-dimensional data is needed to build input models for the calculations. The apex morphology is determined by the crystal faceting \cite{ref20}, which can be easily obtained using the Wulff construction rules \cite{ref21}. For the Pt structures in our case (figure \ref{Pt_tip}) only (111) and (100) facets needed to be considered. 
However, additional factors must be taken into account in order to build an atomically sharp [111] tip. 
For example, the atom at the extreme of the tip is located at the center of a regular triangle (2nd layer), which by itself is located over an hexagonal structure (3rd layer, figure \ref{modelo}(a)). Although in this configuration the atom that finishes the tip is not in the expected crystallographic position, this is the unique way to construct the sharp pyramid. The triangular shape of the 2nd layer fixes the apex [100] facet width to be two-atom-row in size (figure \ref{modelo}(a), named AT2 apex). To generate the twinned apexes, we must consider two crystalline domains; a base and a tip. A twin can be also seen as a 60$^{o}$  rotation of one of the domains along the tip axis; in this way, the (111) (or (100)) facets of the tip become (100) (or (111)) facets at the base (see figures \ref{twin} and \ref{modelo}). Although the domains are rotated, they must have exactly the same cross-section at the twin. Briefly, the faceting pattern is determined by two geometrical constrains: a) an atomically sharp apex imposing that the tip domain must be identical to the equivalent region of the AT2 apex; b) the twin position, because the tip (100) (or (111)) facet width at the twin position will determine the width of the (111) (or (100)) facet at the base. As a consequence, these apexes will not always follow rigorously the Wulff surface balance and also, the total number of atoms in the tips will not be constant (see Table \ref{tab1}).

Based on the procedure discussed above, we generated the 3D structures for the calculations. The initial Pt atomic positions followed a geometrically (or twinned) Pt lattice as schematically shown in figure \ref{modelo}, and then we have carried out fully geometrical optimizations. In Table \ref{tab1} we present the relative energy formation: $E_{struc}-NE_{a}$, where $E_{struc}$ is the energy of the configuration, $N$ is the number of the atoms in the structure, and $E_{a}$ the energy of one isolated atom. The results are relative to the lowest energy configuration (AT4).

Theses results can be explained in terms of faceting and stress release energy generated by the twin defect \cite{ref10,ref21}. AT4 is the favored configuration because it allows, at the same time, one-atom tip and an optimized base surface (figure \ref{modelo}(c)). Moving the defect one or two layers up (AT3 and AT2, respectively) cost a large amount of energy (47.02 and 46.57 eV, respectively). These structures present imperfect faceting but the structural instability comes mainly from the loss of 6 atoms. In contrast, moving the defect one layer down (AT5) changes the structural energy by only 1.35 eV. However, moving one layer further (AT6) requires a large amount of energy (26.53 eV), but considerably less than in the AT2 and AT3 cases. Also, the transition from AT5 to AT6 involves only the loss of 3 atoms.

The AT5 apex represents the first available structure to absorb accumulated stress; this is in excellent agreement with our time resolved HRTEM observations which show that AT5 (figure \ref{Pt_tip}(d)) is formed much before the NC thinning and rupture. As for the possible structural transition from AT4 to AT5 apexes, it is important to note that we are dealing with very small systems with a large ratio surface/volume atoms. The twin can be considered as a rotation between two configurations, this is certainly a very low cost process and probably with a low energy barrier for the size of the structures considered here, as described in the quasi-melting of small Au clusters \cite{ref22}. This interpretation can be better evidenced from video in supplementary material in \cite{video1}.

In order to verify whether the obtained results for Pt could be extrapolated to other metals, we repeated the calculations for Au and Ag, those of most used metals in nanocontacts. The results are displayed in Table \ref{tab1} and, the conclusions are identical, suggesting that these are general features of fcc metallic nanocontacts. However, it must be remarked that generated Pt twins are blocked by a quite high energy barrier \cite{PhysRevB.61.4894, Xu}, much higher than themal energy at room temperature. So, calculations negleting temperature effects, such as the one presented here, can not be considered representative and a reliable discription of room temperature Pt NC experiments. On the other side, energy barrier are much lower for Ag and Au \cite{PhysRevB.61.4894, Xu}; this fact can account for the HRTEM results which show that defects are quickly annhilated at room temperature \cite{ref20,PhysRevB.65.153402}. But, it must be expected that the predicted Ag or Au size limit for twin defect formation should be observed in low temperature experiments

\begin{table}
\begin{ruledtabular}
\begin{tabular}{ccccc}
Twin Position & Number   &   Pt  &   Ag  &  Au  \\
          (n) & of atoms &  (eV) &  (eV) & (eV) \\
       2(AT2) &  99      & 46.67 & 32.06 & 34.49 \\
       3(AT3) &  99      & 47.02 & 32.20 & 34.59 \\
       4(AT4) & 105      &  0.00 &  0.00 &  0.00 \\
       5(AT5) & 105      &  1.35 &  0.51 &  0.36 \\
       6(AT6) & 102      & 26.53 & 17.90 & 18.99 \\
\end{tabular}
\caption{DFT formation energy estimated for twinned Pt, Ag, and Au NCs, with relation to the AT4 tip. See figure 3} \label{tab1}
\end{ruledtabular}
\end{table}

In summary, we have shown that the elongation of Pt NCs induces the formation of twins located a few planes away from the apex, where the atomic arrangement of the atomically sharp Pt pyramid remains close to the ideal one. This means that it is possible to produce one atom nanocontacts with a well defined structure. Many models in molecular electronics have assumed perfect tips, which has been object of criticism as being unrealistic. Our results showed that this is in fact a good approximation. On the other hand, we always will have the presence of extended defects at the apex that would work as a scattering plane, thus it is expected that full electron transmission coefficients are not longer possible, at least for Pt NCs. These aspects should be taken into account when modeling the typical two point electrical measurement on molecular electronic devices. Similar results were also obtained for Au and Ag, suggesting that this is a general behavior of face-centered-cubic metals. Nevertheless they should be only observed at lower temperates (few Kelvin). It seems that 4 and 5 atomic planes is the limit size where twin defects can exist in Pt nanotips, thus representing the frontier to macroscopic plasticity regime for this metal at room temperatures.


This work was supported in part by IMMP/MCT, IN/MCT, THEO-NANO, LNLS, CNPq and FAPESP. The authors acknowledge P.C. Silva and J. Bettini for assistance during the sample preparation and data treatment. 

\bibliography{biblio}

\end{document}